\documentclass[aps,prl,twocolumn,floatfix,amsmath]{revtex4-1} 
\usepackage{graphicx}   
\usepackage{hyperref}
\usepackage{xcolor} 
\newcommand{\unit}[1]{\ensuremath{\, \mathrm{#1}}} 
\begin{document}

\title{Injection-locked diode laser current modulation for Pound-Drever-Hall frequency stabilization using transfer cavities}
\author{C.~E.~Liekhus-Schmaltz,$^1$ R. Mantifel,$^{1,2}$ M. Torabifard,$^{1}$ I.~B.~Burgess,$^{1,3}$ J.~D.~D.~Martin$^{1}$}
\address{$^1$ Department of Physics and Astronomy and Institute for Quantum Computing, \\ University of Waterloo, Waterloo ON, N2L 3G1, Canada}
\address{$^2$ Currently with Department of Physics, 
McGill University, \\ 3600 rue University, Montr\'{e}al QC, H3A 2T8, Canada }
\address{$^3$ Currently with School of Engineering and Applied Sciences,\\ Harvard University, Cambridge, MA, 02138, USA}
\date{\today}\begin{abstract}
A phase modulated RF current source is applied to an injection locked diode laser operating at $780\unit{nm}$.   This produces tunable phase modulated sidebands of the laser suitable for stabilizing the length of an optical transfer cavity using the Pound-Drever-Hall technique.  
The Pound-Drever-Hall signal is anti-symmetric about the lock point, despite the presence of significant diode laser amplitude modulation. 
The stabilized optical transfer cavity is used to frequency stabilize a $776\unit{nm}$ external cavity diode laser.   The stability and tunability of this transfer cavity locked laser is established by observation of the hyperfine components of the $^{87}$Rb $5P_{3/2}-5D_{5/2}$ transition in a vapor cell.
\end{abstract}
\maketitle

Lasers are often frequency stabilized using optical cavity modes.  
Although stable passive cavities can be constructed \cite{ludlow:2007_long}, it is also possible to {\em actively} stabilize a cavity, using a second laser that is frequency stabilized using an atomic reference \cite{burghardt:1979}.  The cavity length can be adjusted with a high bandwidth piezoelectric actuator (PZT) \cite{briles:2010} to keep it in resonance with the reference laser.  Since the cavity {\em transfers} the stability of a reference laser to a target laser it is often referred to as a ``transfer cavity.''  The two lasers may be at very different wavelengths, provided the cavity finesse remains sufficiently high at the two wavelengths.

To stabilize the target laser at an arbitrary frequency, a tunable frequency shift of either the target or reference laser is necessary.  This shift may be obtained by electro-optic modulators \cite{burghardt:1979}, acousto-optic modulators \cite{plusquellic:1996}, or current-modulated injection-locked diode lasers \cite{bohlouli:2006}.  

We demonstrate that a tunable frequency shift and the modulation required for Pound-Drever-Hall (PDH) locking \cite{drever:1983_long} can be obtained by applying a phase modulated RF current to an injection locked slave diode laser.  This modulation produces tunable phase modulated sidebands of the reference laser.  By locking a transfer cavity to one of these sidebands, its length is controlled by a stable, tunable reference frequency, and the frequency of a target laser may be precisely controlled. 

The mechanism for producing sidebands can be intuitively understood from the general behavior that injection locked systems exhibit \cite{siegman:1986}:  the phase difference between the injected and oscillator signal is $\phi=\sin^{-1}(\{\omega_0-\omega_1\}/\omega_m)$, where $\omega_0$ is the free running oscillator angular frequency, $\omega_1$ is the injected signal angular frequency, and $\omega_m$ is the locking half-width (the total locking range being $2\omega_m$).
As the phase difference between the injected and locked oscillator signal depends on the detuning between them, modulation of the locked oscillator resonance frequency will phase modulate its output.  

The lasing frequency of a conventional Fabry-Perot diode laser can be rapidly changed with current.  Kobayashi and Kimura \cite{kobayashi3:1982} demonstrated that by sinusoidally modulating the current applied to an injection locked diode laser, they could observe frequency sidebands corresponding to phase modulation.  In particular, provided the frequency content of the injection current $I(t)=I_{\rm dc}+\Delta I(t)$ is within the locking half-width $\omega_m$, the output of the slave laser is of the form:
\begin{equation}
\tilde{E}=\tilde{E}_0 \exp{\{ j (\omega_0 t + k \Delta I(t) ) \}},
\end{equation}
where $\Delta I(t)$ represents the deviation of the current from the dc value corresponding to $\omega_0=\omega_1$, and $k \approx (\partial \omega_0/\partial I)/\omega_m$.
Sinusoidal modulation of $\Delta I(t)$ produces phase modulation:
$\tilde{E}=\tilde{E}_0 \exp{\{ j (\omega_0 t + \alpha \sin[\delta \: t] ) \}}$, with sidebands in frequency space at $\omega_0 \pm \delta$ with
powers of $P_1=P [J_1(\alpha)]^2$ where $P$ is the total power.  As $\delta \rightarrow \omega_m$ this ceases to be true and the output is more accurately described as being frequency-modulated rather than phase-modulated \cite{kobayashi3:1982}.  This simple picture neglects the
amplitude modulation that must accompany diode laser current modulation.
We will return to this important issue after presentation of the experimental results, which unexpectedly show no adverse effects due to residual amplitude modulation.

Here we consider a current source with phase modulation at an angular frequency $\Omega$ that is expected to produce a slave laser output:
\begin{equation}
\tilde{E}=\tilde{E}_0 \exp{\{ j (\omega_0 t + \alpha \sin[\delta \: t + 
\beta \sin(\Omega t)]) \}}.
\end{equation}
With $\delta \gg \Omega$ this output corresponds to tunable phase modulated sidebands centered around $\omega_0 \pm \delta$.  The cavity may be locked to one of the sidebands using the standard PDH scheme where the modulation frequency is $\Omega$.  The cavity length may then be scanned by varying $\delta$ (keeping $\Omega$ constant).  If a second ``target laser'' is locked to another mode of the same cavity, its frequency may be controlled using $\delta$.

\begin{figure}[tb]
\centering
\includegraphics[scale=.1]{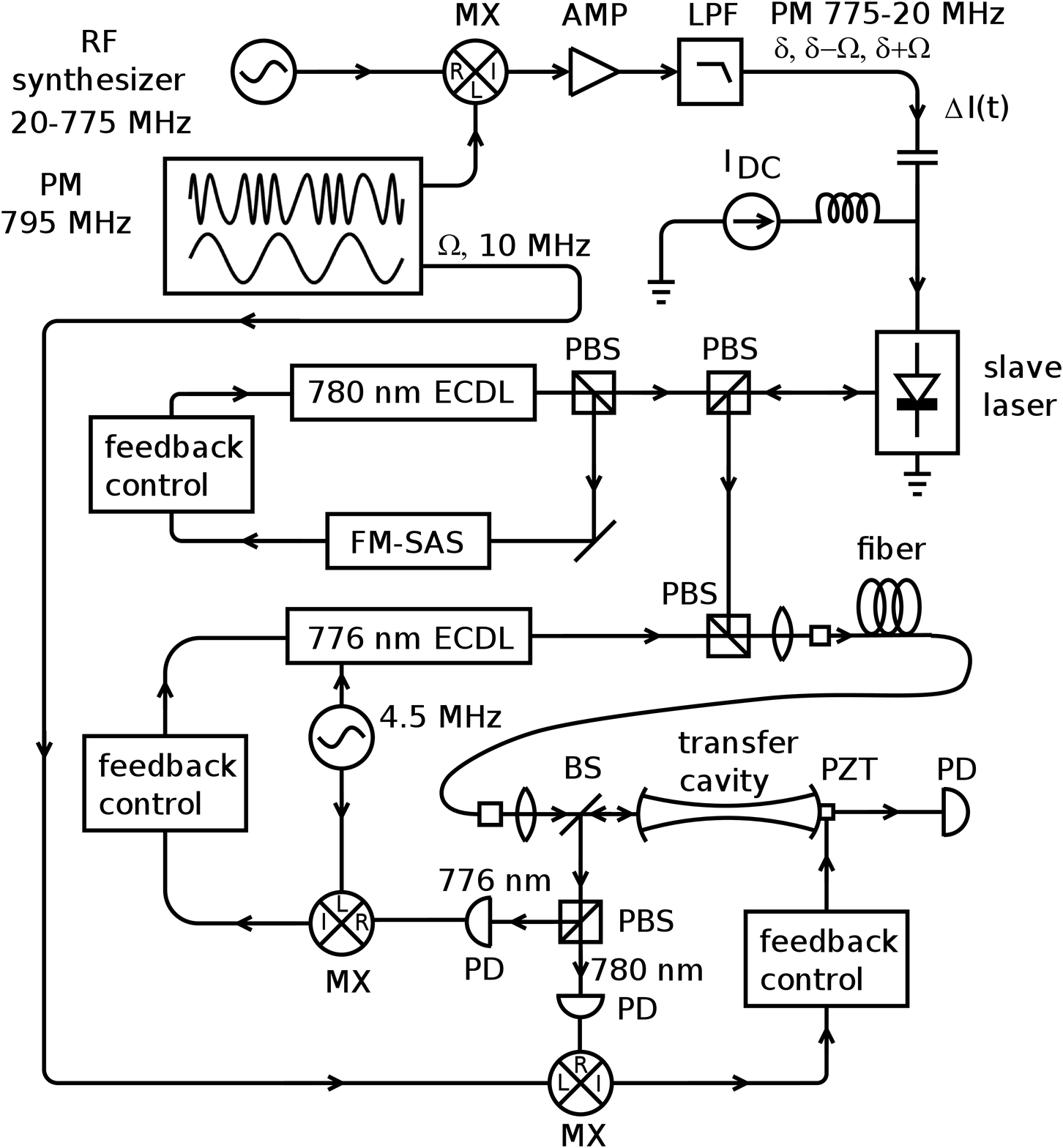}
\caption{
PDH transfer cavity locking of a $776\unit{nm}$ laser using a $780\unit{nm}$ modulated slave laser.  The $776\unit{nm}$ and $780\unit{nm}$ beams have orthogonal polarizations, and are combined and separated using polarizing beam splitter cubes.
AMP: amplifier,
BS: non-polarizing beam splitter,
ECDL: external cavity diode laser, 
FM-SAS: frequency modulated saturated absorption spectroscopy, 
LPF: low-pass filter,
MX: mixer,
PBS: polarizing beam splitter,
PD: photodiode, 
PM: phase modulation,
PZT: piezo-electric transducer.
\label{fig:tpms}
}
\end{figure}

This technique requires a tunable phase modulated current source.   In our implementation, a voltage controlled oscillator (VCO, Minicircuits ZX95-850+) is used to create a phase modulated RF current source, with a fixed center frequency of $795\unit{MHz}$, and a $10\unit{MHz}$ ($\Omega/(2\pi)$) phase modulation frequency.  The VCO output is mixed with a synthesized RF source with a tunable frequency, $\delta$ of 20 MHz to 775 MHz.  The output of the mixer is amplified and low-pass filtered (for rejection of the sum frequency) to obtain a tunable phase modulated current source centered around frequencies from 775 MHz to 20 MHz ($\delta/(2\pi)$).  This signal is applied through a bias-T to an injection locked slave diode laser (Sanyo DL7140-201S) operating at $780\unit{nm}$ (see Fig.~\ref{fig:tpms}).  The master laser is frequency stabilized to a Rb transition in a cell using FM spectroscopy \cite{bjorklund:1980}. 

An RF reference cavity was used to stabilize the center frequency of the modulated VCO.
The $10\unit{MHz}$ phase modulation of the VCO is used to lock its center frequency ($795\unit{MHz}$) to this cavity using the Pound technique \cite{pound:1946} (see Fig.~8 of \cite{cels:2011}).
An integrator varies the input voltage to the VCO based on the Pound error signal.  The reference cavity is a coaxial $\lambda/4$ resonator with silver-plated surfaces, an invar inner conductor and a copper outer conductor and endcaps.  Coupling to the cavity is via a wire loop attached to an SMA connector, threaded into the voltage node endcap.  The loaded and unloaded quality factors of the cavity are 3500 and 4500 respectively.  

Locking the VCO to the cavity significantly improves its long term stability.  Deviations over a 10 hour period were reduced by a factor of 10 to within 5 KHz (observed using a frequency counter with a $1\unit{s}$ gate).  For lower drift, the cavity could be evacuated and temperature stabilized.  Using an RF spectrum analyzer we observe that spurs within $\approx 100\unit{kHz}$ of the carrier are suppressed by locking the VCO and the phase noise of the locked VCO remains lower than the free VCO up to $\approx 2 \unit{kHz}$ from the carrier. We expect that the high frequency phase noise performance could be improved by modifying the simple integrator feedback control. 

We note that this technique (mixing with a phase modulated, frequency stabilized VCO) is also suitable for the generation of tunable phase modulated sidebands using electro-optic modulators, especially broadband fiber-based modulators used for transfer cavity locking (see, for example, \cite{youn:2010}).  In cases where the locked frequency stability is sufficient, it offers an inexpensive alternative to commercial synthesizers.

\begin{figure}[tb]
\centerline{\includegraphics[width=3.375in]{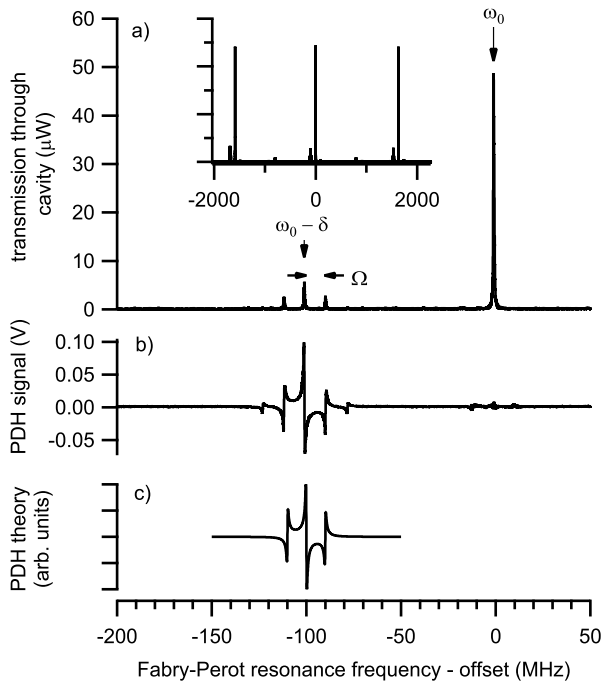}}
\caption{
(a) Fabry-Perot transmission spectra, and (b) PDH error signal for $\delta/(2\pi)=100\unit{MHz}$ observed using a scanning Fabry-Perot cavity with a free spectral range ($FSR$) of $1.6\unit{GHz}$ and finesse (${\cal F}$) of $2500$.  No offset has been added to this signal. (c) Calculated PDH error signal shown for comparison (see, for example \cite{cels:2011}).
\label{fig:fabryperot}
}
\end{figure}

Before applying the phase modulated RF we verify injection locking using a scanning Fabry-Perot cavity.  By simultaneous observation of the $^{85,87}$Rb, $5S_{1/2}-5P_{3/2}$ saturated absorption spectra using both the slave and master lasers we can measure and maximize the locking bandwidth.  To maintain sidebands of approximately equal strength as $\delta$ is varied, it is desirable that the locking bandwidth exceed the maximum modulation frequency ($775\unit{MHz}$ in this case). Using $2\unit{mW}$ of master laser power, we obtain a locking bandwidth of $2\omega_m/(2\pi) \approx 2.8\unit{GHz}$.  

With $10\unit{mW}$ of phase modulated RF incident on the slave laser (measured with a directional coupler), the output of the slave laser develops sidebands at $\pm \delta$ away from the master laser frequency (see Fig.~\ref{fig:fabryperot}a).  These sidebands are phase modulated by $\Omega$, allowing observation of a PDH signal (Fig.~\ref{fig:fabryperot}b).  The PDH signal is obtained by mixing the photodiode current obtained from reflection off the front of the cavity with the 10 MHz used to phase modulate the VCO (with an appropriate phase shift applied).

\begin{figure}[tb]
\centerline{\includegraphics[scale=.1]{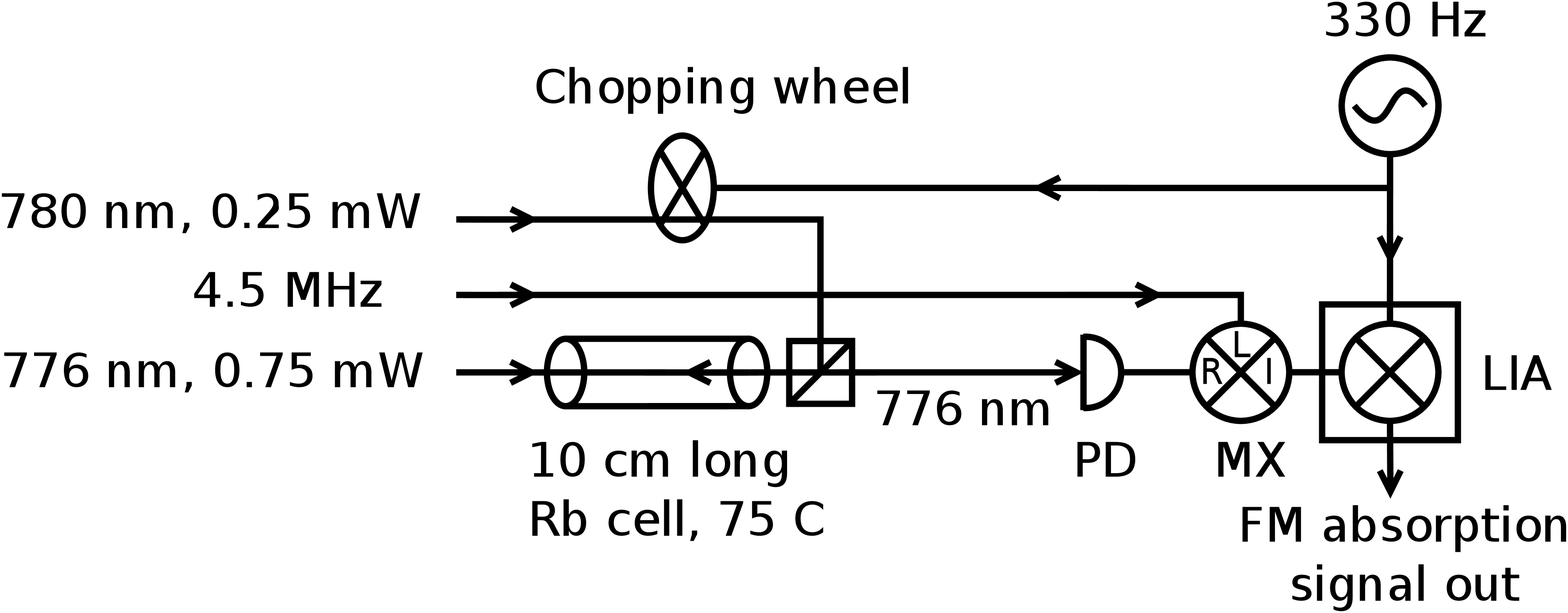}}
\caption{
Scheme to monitor $776\unit{nm}$ ECDL tunability and frequency stability
using the $^{87}$Rb, $5P_{3/2}-5D_{5/2}$ transitions.
LIA: lock-in amplifier
MX: mixer,
PD: photodiode
\label{fig:abssetup}
}
\end{figure}

A transfer cavity could be locked to either one of the $\omega_0 \pm \delta$ sidebands. We find that they differ in amplitude and that the $\omega_0+\delta$ sideband has an amplitude that is more susceptible to the exact setting of the dc current of the injection locked laser.  A relatively long PZT travel allows a choice of cavity modes so that the more stable $\omega_0-\delta$ sideband can always be used.
We note that the difference in the behavior of the sidebands is inconsistent with our simple phase modulation picture which neglects the amplitude modulation that must accompany current modulation.  

In view of the large difference between the magnitudes of the $\omega_0 \pm \delta$ sidebands (the $\omega_0+\delta$ sideband is barely visible in the inset to Fig.~\ref{fig:fabryperot}) it is perhaps surprising that the $\omega_0-\delta+\Omega$ and $\omega_0-\delta-\Omega$ sidebands have the same magnitude.  It is often observed that direct current modulation of a diode laser with a {\em single tone} at $\Omega$ produces asymmetric sidebands, and thus imperfect PDH signals.
However symmetric {\em secondary} sideband magnitudes are consistent with a straightforward model of the modulated slave laser output that includes amplitude modulation.

To rationalize why the secondary sideband magnitudes are equal, consider a model for the output of the slave laser that includes amplitude modulation:
\begin{equation}
\frac{\tilde{E}}{\tilde{E_0}}=(1+A\sin[\phi]+B\cos[\phi])\exp \{j(\omega_0 t+\alpha \sin[\phi]) \},
\label{eq:ammod}
\end{equation}
where $\phi=\delta t +\beta \sin(\Omega t)$.
For relatively low $\delta$ we might expect that the phase and amplitude modulation would be in phase and $B=0$.  However, this is inconsistent
with our observation of asymmetric sideband amplitudes at $\omega_0+\delta$ and $\omega_0-\delta$ when we apply a single tone at $\delta$ (see, for example, Table II of \cite{kobayashi:1982}).  
There must be simultaneous amplitude and phase modulation, and a phase shift between them.  This phase shift may be calculated from a detailed knowledge of the laser and the operating parameters, as Lidoyne {\it et al.}~\cite{lidoyne:1991} have shown.

Using the Jacobi-Anger identity to expand Eq.~\eqref{eq:ammod} we find that the amplitude of the $\omega_0-\delta+\Omega$ frequency component is
\begin{equation}
\begin{split}
	\frac{\tilde{E}_{\omega_0-\delta+\Omega}
}{\tilde{E_0}} = & J_1(\alpha)J_1(\beta)
-\frac{(Aj+B)}{2}J_1(\beta)J_0(\alpha)\\
	&
+\frac{(-Aj+B)}{2}J_2(\alpha)\sum_{\ell=-\infty}^{\infty}J_{\ell}(\beta)J_{\ell-1}(2\beta),
\end{split}
\end{equation}
where $J_{\ell}$ is a Bessel function of order $\ell$.
The $\omega_0-\delta-\Omega$ frequency component has the same magnitude but is of opposite sign.  Thus these sidebands are suitable for the PDH technique \cite{drever:1983_long,cels:2011}, which is consistent with our observations.  A similar result is found for the $\omega_0+\delta+\Omega$ and $\omega_0+\delta-\Omega$ frequency components.

\begin{figure}[tb]
\centerline{\includegraphics{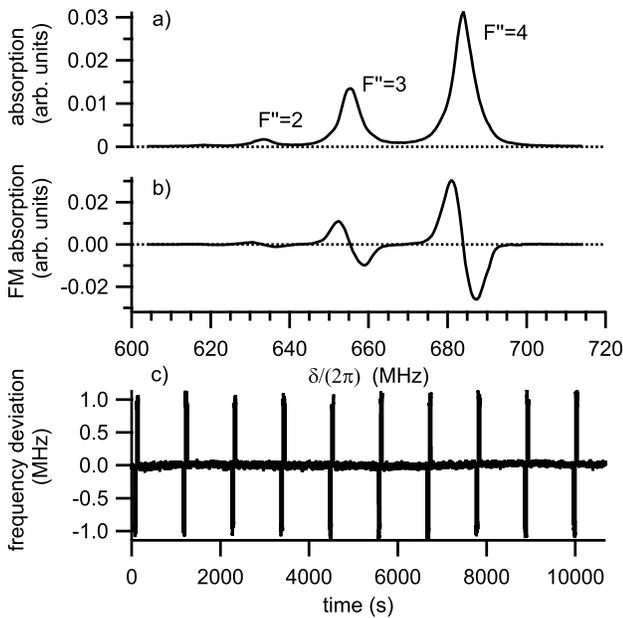}}
\caption{
(a) Absorption spectrum of the $5P_{3/2}-5D_{5/2}$ transition of $^{87}$Rb
(b) FM absorption spectrum (see Fig.~\ref{fig:abssetup})
(c) Frequency stability of the locked $776\unit{nm}$ laser as a function of time, monitored using the FM absorption signal ($0.3\unit{s}$ time constant).  The RF frequency is set to place the $776\unit{nm}$ laser at the $5P_{3/2},F'=3\rightarrow 5D_{5/2},F''=4$  
transition and the FM signal is used as a frequency discriminator.  Periodically stepping $\delta/(2\pi)$ up and down by $1\unit{MHz}$ allows conversion of the FM signal to frequency deviation.
\label{fig:775absres}
}
\end{figure}

To demonstrate the tunability of a transfer cavity locked using the modulated injection locked slave laser, we have stabilized the frequency of a $776 \unit{nm}$ external cavity diode laser.  The transfer cavity had a $FSR=1.9\unit{GHz}$ and  ${\cal F}=350$.   Its length is locked using a PM tunable sideband of the slave laser, as shown in Fig.~\ref{fig:tpms}.  The $776\unit{nm}$ diode laser is then locked to a cavity resonance using PDH locking.  This laser is current modulated at $4.5\unit{MHz}$ to provide PM sidebands suitable for locking.

The $^{87}$Rb $5P_{3/2}-5D_{5/2}$ transition \cite{grove:1995_long} is used to observe the locked $776\unit{nm}$ laser's tunability and frequency stability.  A $780\unit{nm}$ counter-propagating beam is sent through a Rb vapor cell exciting the $5P_{3/2},F'=3$ states, and the $776\unit{nm}$ absorption measured with a photodiode (Fig.~\ref{fig:abssetup}).  We demodulate the absorption using the $4.5\unit{MHz}$ modulating source to produce an FM, dispersion-like absorption signal \cite{bjorklund:1980}.  To improve signal to noise, the $780\unit{nm}$ beam intensity is modulated with a chopping wheel, and both the absorption and FM signals detected with a lock-in amplifier.
By varying $\delta$ we may scan the frequency of the $776\unit{nm}$ laser over the absorption lines (see Fig.~\ref{fig:775absres}a).  

As with Grove {\it et al.}~\cite{grove:1995_long} we can compared the observed frequency differences between absorption lines with the results of 
Nez et al.~\cite{Nez1993432}.  The correspondence between  a frequency change in the locked laser $\Delta f_{776}$ and a frequency change in the tunable sideband $\Delta f_{780}$ is: $\Delta f_{776}/\Delta f_{780} \approx f_{776}/f_{780} \approx 1.0055$.  Using this factor, we determine from the absorption spectrum in Fig.~\ref{fig:775absres}a) that the energy level difference between $5D_{5/2}$ $F''=4$ and $F''=3$ is $28.82 \pm 0.03 \unit{MHz}$, 
and between $F''=3$ and $F''=2$ it is $22.7 \pm 0.3 \unit{MHz}$.
These results are consistent with $28.82 \pm 0.01\unit{MHz}$ and 
$22.96\pm0.01\unit{MHz}$ from \cite{Nez1993432}.

As $\delta$ is varied the dc offset of the PDH signal varies, possibly due to amplitude modulation of the slave laser at $10 \unit{MHz}$. Although the magnitude of this offset is typically small compared to the peak PDH signal,
this varying offset can introduce non-linearities in the scanned laser frequency.  In general this effect is reduced for larger $\delta$.  For example, by recording PDH spectra as a function of $\delta$ for the cavity of Fig.~\ref{fig:fabryperot} we have established an upper bound of $50\unit{kHz}$
on scan non-linearities, and a variation in the PDH signal magnitude of less than 50\% for $\delta=500\unit{MHz}$ to $\delta=750\unit{MHz}$.  Although we have not done so, the influence of the dc offset can be reduced by deriving the PDH signal from the difference of two photodiode signals: one which measures the light incident on the cavity, and the other which measures the light reflected from the cavity.  The relative powers incident on each photodiode should be adjusted to null the PDH error signal when the cavity is not resonant with the incident light.

The long-term frequency fluctuations of the $776\unit{nm}$ frequency stabilized laser may be monitored using the $5P_{3/2}-5D_{5/2}$ FM signal.   By setting
$\delta$ to the strongest absorption line ($F''=4$), the FM signal (Fig.~\ref{fig:775absres}b) may be used to provide an upper bound on the locked laser frequency fluctuations (Fig.~\ref{fig:775absres}c).
The observed frequency stability is sufficient for many laser locking applications.  For example, this system has been in regular use for several months, to frequency stabilize a Ti:sapphire laser operating at $960\unit{nm}$ for the excitation of laser cooled atoms to Rydberg states \cite{petrus:2008}. The lock is typically maintained over several hours, usually limited by the need to relock the master laser.

In summary, we have demonstrated that with the appropriate current modulation, tunable phase modulated sidebands can be put on injection locked diode lasers.  These sidebands may be used to stabilize and vary the lengths of PZT based optical transfer cavities using the PDH technique.  

We thank L. Jones for providing the heated Rb cell.  This work was supported by NSERC.


\end{document}